\documentclass{article}

\usepackage{amssymb,amsfonts,amsmath,stmaryrd}
\usepackage{cite,enumerate,float,indentfirst}
\usepackage{color}
\usepackage{tikz}
\usetikzlibrary{arrows,snakes,backgrounds}
\usepackage{ytableau}
\usepackage[vcentermath]{youngtab}
	\definecolor{almond}{rgb}{0.94, 0.87, 0.8}
\usepackage{xcolor}

\def\be{\begin{eqnarray}}
\def\ee{\end{eqnarray}}

\newcommand{\ct}{\tilde{c}}

\def\p{\partial}
\def\tr{{\rm tr}\,}

\definecolor{red}{rgb}{1,0,0}
\definecolor{orange}{rgb}{1,0.5,0}
\definecolor{violet}{rgb}{0.7,0,1}

\hoffset= - 1.0in         % switch off for draft style
\begin{document}

\begin{center}
\begin{small}
\hfill FIAN/TD-04/21\\
\hfill IITP/TH-07/21\\
\hfill ITEP/TH-10/21\\
\hfill MIPT/TH-06/21\\
\end{small}
\end{center}

\vspace{.5cm}

\begin{center}
\begin{Large}\fontfamily{cmss}
\fontsize{17pt}{27pt}
\selectfont
	\textbf{Virasoro versus superintegrability. Gaussian Hermitian model}
	\end{Large}
	
\bigskip \bigskip
%\hspace{-1cm}
\begin{large}A. Mironov$^{a,b,c,}$\footnote{mironov@lpi.ru; mironov@itep.ru},
V. Mishnyakov$^{d,a,b,}$\footnote{mishnyakovvv@gmial.com},
A. Morozov$^{d,b,c,}$\footnote{morozov@itep.ru},
R. Rashkov$^{e,f,}$\footnote{rash@phys.uni-sofia.bg; rash@hep.itp.tuwien.ac.at}
 \end{large}
\\
\bigskip

\begin{small}
$^a$ {\it Lebedev Physics Institute, Moscow 119991, Russia}\\
$^b$ {\it ITEP, Moscow 117218, Russia}\\
$^c$ {\it Institute for Information Transmission Problems, Moscow 127994, Russia}\\
$^d$ {\it MIPT, Dolgoprudny, 141701, Russia}\\
$^e$ {\it Department of Physics, Sofia University,
	5 J. Bourchier Blvd., 1164 Sofia, Bulgaria} \\
$^f$ {\it ITP, Vienna University of Technology,
	Wiedner Hauptstr. 8–10, 1040 Vienna, Austria   }
\end{small}
 \end{center}
\medskip

\begin{abstract}
Relation between the Virasoro constraints and KP integrability (determinant formulas)
for matrix models is a lasting mystery.
We elaborate on the claim that the situation is improved when integrability
is enhanced to super-integrability, i.e. to explicit formulas for
Gaussian averages of characters.
In this case, the Virasoro constraints are equivalent to simple recursive formulas,
which have appropriate combinations of characters as their solutions.
Moreover, one can easily separate dependence on the size of matrix, and deduce superintegrability
from the Virasoro constraints.
We describe one of the ways to do so for the Gaussian Hermitian matrix model.
The result is a spectacularly elegant reformulation of Virasoro constraints
as identities for the Schur functions
evaluated at appropriate loci in the space of time-variables.
\end{abstract}

\bigskip

\bigskip

One of the basic results in the theory of matrix models \cite{UFN31}-\cite{UFN36}
is that they satisfy a set of Ward identities, which allow one to unambiguously determine their
Gaussian partition functions,
%at least in the perturbative Gaussian phase.
and specify their non-perturbative continuations beyond the Gaussian phase.
Mostly publicized (though actually not the simplest) example is the
Gaussian Hermitian model \cite{Dyson1,Dyson2},
which is associated with the partition function
\be
{\cal Z}\{p\} = \int_{N\times N} dM e^{-\tr M^2} e^{\sum_k {p_k\over k} \tr M^k}
\ee
This partition function is understood as a (graded) power series in $p_k$, and normalized so that ${\cal Z}\{0\}=1$.
As a corollary of invariance of the integral,
it satisfies the Virasoro constrains \cite{Vir1}-\cite{Vir4}
\be\label{Vir0}
\hat L_n {\cal Z}\{p\} = 0
\ee
with
\be\label{Vir}
\hat { L}_n := \sum_{k} (k+n)p_k\frac{\p}{\p p_{k+n}} + \sum_{a=1}^{n-1} a(n-a)\frac{\p^2}{\p p_a\p p_{n-a}}
+ 2Nn\frac{\p}{\p p_n}    + N^2\delta_{n,0} + Np_1 \delta_{n+1,0}
- (n+2)\frac{\p}{\p p_{n+2}}
\ee
One of the essential properties of the model is that these Virasoro constraints have a unique
power series solution.

A more recent observation \cite{MM1}-\cite{MM3} (see also \cite{Orlov}) is that this model is also {\it super}\,integrable,
i.e. the coefficients $c_R$ of the expansion
\be\label{Zchi}
{\cal Z}\{N,p\} = \sum_R c_R(N)\cdot \chi_R\{p\}
\ee
in the Schur polynomials $\chi_R\{p\}$ are also exactly calculable and
expressed in terms of the same $\chi_R$ evaluated at some special loci.
In particular,
\be\label{cR}
c_R(N) = \ct_R\cdot {\chi_R\{N\}\over\chi_R\{\delta_{k,1}\}}=\ct_R \cdot \prod_{\Box \in R} (N+j_\Box-i_\Box)
\ee
where $\ct_R$ do not depend on $N$ and are actually equal to
\be\label{cR2}
\ct_R = \chi_R\{\delta_{k,2}\}
\ee

In this letter, we explain that this description can be, in fact, derived
from the Virasoro constraints,
which contributes to our understanding of the triality
superintegrability-Ward identities-integrability,
and can be further generalized to a variety of other eigenvalue, matrix and tensor models.
In fact, this can be done in many ways. An example of straightforward
and detailed calculation in a more sophisticated model of \cite{Qfns} can be found in \cite{China1,China2}.
We are going to review some of these ways elsewhere,
and, in this letter, we concentrate on an elegant and non-evident bypass,
which seems to maximally exploit the beauty of superintegrability.

\bigskip

Our logic is as follows.
\begin{itemize}
\item[1)] We start with the expansion (\ref{Zchi}) and demonstrate that, in application to this ansatz,
the set of all Virasoro constraints reduces to a set of two equations
\begin{equation}\label{virc}
\boxed{
\ {\rm Virasoro} \ \ \ \ \ \ \ \ \Longleftrightarrow \ \ \ \ \ \ \ \
    \begin{split}
         & \sum_{R+\Box} c_{R+\Box}(N) = \sum_{R-\Box} (N+ j_\Box-i_\Box)\cdot c_{R-\Box}(N)\\
          &  \sum_{R+\Box} (j_\Box-i_\Box) \cdot c_{R+\Box}(N) =
          \sum_{R-\Box} (N+ j_\Box-i_\Box)^2 \cdot c_{R-\Box}(N)
    \end{split} \
    }
\end{equation}
where $(i_\Box,j_\Box)$ are the coordinates of the square removed from or added to the Young diagram $R$.
\item[2)] We explain that the system of two equations of the form
\be
\sum_{R+\Box} \alpha_m(\Box)\cdot c_{R+\Box}(N) =
\sum_{R-\Box} \beta_m(\Box) \cdot c_{R-\Box}(N),\ \ \ \ \ \ \ m=1,2
\ee
in the case of non-degenerate coefficients $\alpha_m$, $\beta_m$ always have a unique solution.
\item[3)]
One could just use the answer, but a more elegant option is to separate the $N$  dependence by (\ref{cR}),
and then reduce the equations (\ref{virc}) to a new system of equations that fix the
$N$-independent $\ct_R$:
%Using the form (\ref{cR}), we show that (\ref{virc}) are equivalent to the three equations for $\ct_R$
\begin{equation}\label{virct}
\boxed{
\ {\rm Virasoro} \ \ \ \ \ \ \ \ \Longleftrightarrow \ \ \ \ \ \ \ \
\begin{split}
        &\sum_{R+\Box} \ct_{R+\Box}= 0\\
        &\sum_{R+\Box}(j_\Box-i_\Box)\cdot \ct_{R+\Box}=\sum_{R-\Box} \ct_{R-\Box} \\
        &\sum_{R+\Box}(j_\Box-i_\Box)^2 \cdot \ct_{R+\Box}=\sum_{R-\Box} (j_\Box-i_\Box)\cdot \ct_{R-\Box}
\end{split} \
}
\end{equation}
Any two of these three equations already have a unique solution in accordance with 2).
Hence, the system is overfull,
and, finding its solution, one makes an additional check that the decomposition (\ref{cR}) is, indeed, correct.
%and one has to check that the form (\ref{cR}) is, indeed, correct.
\item[4)] At last, we demonstrate that, indeed, $\tilde c_R$ as in (\ref{cR2}) solves all these three equations, and, hence, solves the Virasoro constraints.
\end{itemize}

Our consideration is based on a set of combinatorial identities, which can be obtained either from the fermionic representation of the Schur functions \cite{DJKM}, or by a direct computation using the Littlewood-Richardson coefficients. This later approach is briefly explained in a short Appendix.
Still, the spectacularly simple form of the equations (\ref{virc})  and (\ref{virct})
calls for alternative interpretations, which could bypass the details of representation theory
and make the relation between superintegrability
and Ward identities more straightforward and functorial.

\paragraph{1.}  We split all system of the Virasoro constraints (\ref{Vir0})-(\ref{Vir}) into two parts
\begin{equation}\label{vir2}
\begin{split}
        &\hat{L}_{-1} {\cal Z}\{N,p\} =0
    \\& \sum (k-1) p_k \hat{L}_{k-1} {\cal Z}\{N,p\}=0
\end{split}
\end{equation}
Consider first the action of the $\hat{L}_{-1}$-constraint. Using particular cases of formulas (\ref{der1}), (\ref{derk}), one obtains
\be\label{Lm1}
\hat{L}_{-1}\chi_R=\sum_{\Box}(N+j_\Box-i_\Box)\chi_{R+\Box}-\sum_\Box\chi_{R-\Box}
\ee
Inserting this into (\ref{Zchi}), one finally obtains
\be
\sum_{R+\Box} c_{R+\Box}(N) = \sum_{R-\Box} (N+ j_\Box-i_\Box) c_{R-\Box}(N)
\ee
Now using the notation for the ``classical" part of the $W$-operator
\begin{equation}
  \hat  w_{n}:=\sum p_k \hat l_{n+k},\ \ \ \ \ \
\hat l_n:=\sum (k+n) p_{k} \frac{\partial}{\partial p_{k+n}}+\sum_{a=1}^{n-1}a(n-a) \frac{\partial^{2}}{\partial p_{a} \partial p_{n-a}}
\end{equation}
one can rewrite the second equation of (\ref{vir2}) as
\begin{equation}\label{wn}
    \left(\hat w_n + 2N \sum (k+n)p_k \dfrac{\partial }{\partial p_{n+k}} +N^2 p_{-n}- \sum (n+2+k)p_k \dfrac{\partial }{\partial p_{n+2+k}} \right){\cal Z}\{N,p\} =0
\end{equation}
In fact, it is enough to consider only the lowest equation of the infinite system (\ref{wn}), that at $n=-1$ in order to unambiguously fix ${\cal Z}\{N,p\}$. That is, using that
\begin{equation}\label{wm1}
    \hat w_{-1} \chi_R = \sum_{R+\Box} (j_\Box-i_\Box)^2 \chi_{R+\Box}
\end{equation}
one immediately produces for the coefficients of expansion (\ref{Zchi}) the equation
\begin{equation}
    \sum_{R+\Box} (j_\Box-i_\Box) c_{R+\Box}(N) = \sum_{R-\Box} (N+ j_\Box-i_\Box)^2 c_{R-\Box}(N)
\end{equation}

\paragraph{2.}
The next step is to explain why our procedure with replacing the whole Borel part of the Virasoro algebra
by the $L_{-1}$- and $w_{-1}$-constraints leads to a unique solution, which can be constructed by solving equations (\ref{virc}) recursively.
In fact, an even a stronger claim is correct: two equations of the form
\be
\sum_{R+\Box} \alpha_m(\Box)c_{R+\Box} = \sum_{R-\Box} \beta_m(\Box) c_{R-\Box},\ \ \ \ \ \ \ m=1,2
\ee
in the case of non-degenerate coefficients $\alpha_m$, $\beta_m$ can be solved recursively to give rise to a unique solution.

\bigskip

Indeed, for $R$ of a given length $l(R)$, these equations introduce only one diagram which is not of length $l(R)$ or smaller, which is  $[R_1,\ldots R_{l(R)},1]$. Because we have two equations, we can exclude this coefficient, and get a recursion involving only diagrams of a fixed length.
At the same time, the equations determine the $c_{[R_1,\ldots R_{l(R)},1]}$ itself, which serves as an initial condition on the recursion at the next level.
\\\\
For example, at level $l(R)=2$, if one goes from $c_{[r,4]}$ to $c_{[r,5]}$,
the equations involve the following partitions:
\begin{center}
   \begin{tikzpicture}
      \ytableausetup{boxsize=0.7em,aligntableaux = center}

      \node (A) at (0,0) {\ydiagram{5,4} };
       \node (B1) at (-2,-1) {
          \ydiagram{5,4}*[*(almond) ]{5+1}
       };
         \node (B2) at (-2,-2) {
          \ydiagram{5,4}*[*(almond) ]{5+0,4+1}
       };
        \node (B3) at (-2,-3) {
          \ydiagram{5,4}*[*(almond) ]{5+0,4+0,1}
       };
        \node (C1) at (2,-1.5) {\ydiagram{4,4} };
        \node (C2) at (2,-2.5) {\ydiagram{5,3} };

        \draw (A)--(B1);
        \draw (A)--(C1);
\end{tikzpicture}
\end{center}
Let us demonstrate how this works order by order, and start from symmetric representations
\begin{equation}
  \left\{  \begin{split}
        &c_{[r+1]}+c_{[r,1]}=(N+r-1) c_{[r-1]}\\
        &r c_{[r+1]}-c_{[r,1]}=(N+r-1)^2 c_{[r-1]}
    \end{split} \right. \  \Rightarrow (r+1)c_{[r+1]} = (N+r-1)(N+r)c_{[r-1]}
\end{equation}
To fix the initial conditions, we choose $R=[\ ]$, which gives
\begin{equation}
c_{[1]}=0
\end{equation}
and we normalize $c_{[ \ ]}=1$.
Hence, one gets,
\begin{equation}
c_{[r]}= \left\{
\begin{array}{cl}
    \dfrac{\prod\limits_{i=1}^{r-1} (N+r-i)}{r!!}   &\ \ \ \ \ \ r =2k
    \\
    &\\
    0   &\ \ \ \ \ \ r =2k+1
\end{array} \right.
\end{equation}
From the same equations, one gets
\begin{equation}
 c_{[r,1]}= \left\{
\begin{array}{cl}
    \dfrac{(N-1)\prod\limits_{k=0}^{r-1} (N+k)}{(r+1)!!}   &\ \ \ \ \ \ r =2k-1
    \\
    &\\
    0   &\ \ \ \ \ \ r =2k
\end{array} \right.
\end{equation}
For partitions of length $l(R)=2$, the recursion starts at
$[r,2]$:
\begin{equation}
\begin{split}
    &\left\{
    \begin{split}
        & c_{[r,1]}+c_{[r-1,2]}+c_{[r-1,1,1]}=(N+r-2)c_{[r-2,1]}+(N-1)c_{[r-1]}\\
        & (r-1)c_{[r,1]}+ \, 0 \,  - 2c_{[r-1,1,1]}=(N+r-2)^2c_{[r-2,1]}+(N-1)^2c_{[r-1]}
    \end{split}
    \right. \Rightarrow \\\\
    &
     \Rightarrow  (r+1)c_{[r,1]}+2c_{[r-1,2]}=(N+r-2)(N+r)c_{[r-2,1]}+(N-1)(N+1)c_{[r-1]}
\end{split}
\end{equation}
Hence,
\begin{equation}
\begin{split}
    &c_{[r,2]}=\dfrac{N(N-1)\prod\limits_{k=0}^{r-1} (N+k)}{2(r!!)}\\
    &c_{[r,1,1]}=\dfrac{r}{2((r+2)!!)} \ \cdot (N-2)(N-1)\prod\limits_{k=0}^{r-1} (N+k)
\end{split}
\end{equation}
for partitions with even $r$, while for those with odd $r$ clearly vanish.
\\\\
For $[r,3]$:
\begin{equation}
    (r+2)c_{[r+1,2]}+3c_{[r,3]}=(N+r-2)(N+2)c_{[r-1,2]}+N^2 c_{[r-1]}
\end{equation}
and so on.

\paragraph{3.} One can find that the solution is of the form (\ref{cR})-(\ref{cR2})
by applying the recursive procedure of the previous paragraph.
Instead here we demonstrate by direct check that the answer is given by (\ref{cR})-(\ref{cR2}).

We start with the form (\ref{cR}). Then,
\be
c_{R+\Box}(N)=\ct_{R+\Box}\cdot (N+ j_\Box-i_\Box)\cdot{c_R(N)\over \ct_R},\ \ \ \ \ \ \
(N+ j_\Box-i_\Box)\cdot c_{R-\Box}(N)=\ct_{R-\Box}\cdot {c_R(N)\over \ct_R}
\ee
Now we insert these formulas into our equations (\ref{virc}),
\be
\sum_\Box (N+j_\Box - i_\Box)\cdot \tilde c_{R+\Box}=\sum_{\Box}\tilde c_{R-\Box},\ \ \ \ \ \
\sum_\Box (j_\Box - i_\Box)\cdot\tilde c_{R+\Box}=\sum_{\Box}(N+j_\Box - i_\Box)\cdot\tilde c_{R-\Box}
\ee
and consider each order of $N$. This produces the {\it three} equations (\ref{virct}).
As we explained in the previous paragraph,
any two of these three equations fix a unique solution and, hence, the system is overfull.
We check in the next paragraph that the solution is (\ref{cR2}),
and it satisfies all the three equations at once.

\paragraph{4.} In order to prove that (\ref{virct}) are satisfied by
\begin{equation}
    \ct_R= \chi_R\{\delta_{k,2}\}
\end{equation}
we use the identities (\ref{wm1}), (\ref{Pieri}), (\ref{der1}), (\ref{derk}),
\begin{equation}\label{onchi}
\begin{split}
        &p_1 \chi_R = \sum_{R+\Box} \chi_{R+\Box}\\
        &\hat l_{-1}\chi_R=\sum_{R+\Box} (j_\Box-i_\Box) \chi_{R+\Box}\qquad \dfrac{\partial  \chi_R}{\partial p_1} = \sum_{R-\Box} \chi_{R-\Box}
        \\
        &\hat w_{-1}\chi_R=\sum_{R+\Box} (j_\Box-i_\Box)^2 \chi_{R+\Box}\qquad \hat l_{1}\chi_R=\sum_{R-\Box} (j_\Box-i_\Box) \chi_{R-\Box}
\end{split}
\end{equation}
and evaluate these identities at the locus $p_k=\delta_{k,2}$:
\begin{equation}
\begin{split}
    &\left. p_1 \chi_R \right|_{p_k=\delta_{k,2}} = 0\\
    &\left. \hat l_{-1} \chi_R \right|_{p_k=\delta_{k,2}}= \left. \sum_k (k-1) p_k \dfrac{\partial \chi_R}{\partial p_{k-1}}  \right|_{p_k=\delta_{k,2}} = \left. \dfrac{\partial \chi_R}{\partial p_{1}}\right|_{p_k=\delta_{k,2}}
    \\
    &\left. \hat w_{-1}\chi_R \right|_{p_k=\delta_{k,2}}= \left. \sum p_k \hat l_{k-1} \chi_R \right|_{p_k=\delta_{k,2}} = \left. \hat l_{1} \chi_R \right|_{p_k=\delta_{k,2}}
\end{split}
\end{equation}
Plugging \eqref{onchi} in here, one gets exactly \eqref{virct} for $\ct_R = \chi_R\{\delta_{k,2} \}$.

\paragraph{Appendix.}
In above considerations,
one can use the following properties of the Schur polynomials \cite{Fulton}:
\begin{itemize}
\item
Pieri's formula,
\be\label{Pieri}
p_1\chi_R\{p\}=\sum_\Box\chi_{R+\Box}\{p\}
\ee
where the sum runs over all possible Young diagrams obtained by adding a square to $R$.
\item Using the Fr\"obenius formula
\be\label{Fro}
p_k=\sum_Q\psi_Q([k])\chi_Q
\ee
where the symmetric group character $\psi_Q([k])=(-1)^a$ for $Q=[a,1^{k-a}]$, and $\psi_Q([k])=0$ otherwise,
and
\be
\chi_Q\Big\{k{\partial\over\partial p_k}\Big\}\cdot\chi_R\{p_k\}=\chi_{R/Q}\{p_k\}
\ee
where $\chi_{R/Q}$ is the skew Schur function, one obtains
\be\label{der1}
k{\partial \chi_R\over\partial p_k}=\sum_Q\psi_Q([k])\chi_{R/Q}=\sum_{a,P}(-1)^a{\cal N}^R_{P,[a,1^{k-a}]}\chi_P
\ee
The Littlewood-Richardson coefficients are defined here by the formula
\be\label{LR}
\chi_P\cdot\chi_Q=\sum_R{\cal N}^R_{PQ}\chi_R
\ee
\item From (\ref{Fro}) and (\ref{LR}), it follows
\be
p_k\chi_R\{p\}=\sum \psi_Q([k]){\cal N}_{QR}^P\chi_P=\sum_{a,P}(-1)^a{\cal N}^P_{R,[a,1^{k-a}]}\chi_P
\ee
and, hence,
\be\label{derk}
\sum kp_{k+n}\frac{\p \chi_R}{\p p_k}=\sum_{S} \phantom{.}_{n}B^{S}_R\chi_S
\ee
where, for instance,
\be
_{\pm 1}B^{S}_{R}:=\sum_{k,a,b,P}(-1)^{a+b}{\cal N}^R_{P,[a,1^{k-a}]}{\cal N}^S_{P,[b,1^{k-b\pm 1}]}=
(j_\Box-i_\Box)\delta_{S,R\pm\Box}
\ee
\end{itemize}

\section*{Acknowledgements}

We are grateful to Y. Zenkevich for a useful discussion. Our work was supported in part by RFBR and NSFB according to the research project number 19-51-18006 (A.Mir., A.Mor.), by RFBR and TUBITAK, project number 21-51-46010 (A.Mir., A.Mor.), by RFBR and MOST, project number 21-52-52004 (A.Mir., A.Mor., V.Mish.). The work was also partly funded by the grant of the Foundation for the Advancement of Theoretical Physics ``BASIS" (A.Mir.), by RFBR grants 19-01-00680 (A.Mir., V.Mish.) and 19-02-00815 (A.Mor.).
R~.R. was supported in part by FNI/BG-RU-2018/246, BNSF Grants H-28/5 and DN-18/1.

\end{document}